\documentclass[12 points]{article}
\usepackage{amssymb}
\usepackage{eucal}
\usepackage{amsmath}
\usepackage{amsthm}
\usepackage{graphicx}
\usepackage{float}

\newcommand{\R}{{\mathbb  R}}
\newcommand{\wt}{\widetilde}
  \numberwithin{equation}{section} \newtheorem{thm}{\bf
Theorem}[section]
  
  \theoremstyle{remark}

\begin{document}

\title{\large\bf A note on stability of nongeneric equilibria for an underwater vehicle} \author{Petre Birtea and Dan Com\u{a}nescu \\ {\small Department of Mathematics, West University of Timi\c soara}\\ {\small Bd. V.
P\^ arvan,
No 4, 300223 Timi\c soara, Rom\^ania}\\ {\small birtea@math.uvt.ro, comanescu@math.uvt.ro}\\ }
\date{ } \maketitle

\begin{abstract}
We study the Lyapunov stability of a family of nongeneric equilibria with spin for underwater vehicles with noncoincident centers. The nongeneric equilibria belong to singular symplectic leaves that are not characterized as a preimage o a regular value of the Casimir functions. We find an invariant submanifold such that the nongeneric equilibria belong to a preimage of a regular value that involves sub-Casimir functions. We obtain results for nonlinear stability on this invariant submanifold. 

\end{abstract}

{\bf MSC}: 53Bxx, 58A05,  34K20, 70Exx, 70Hxx.

{\bf Keywords}: Nonlinear stability, Nongeneric equilibria, Underwater vehicle, Hamilton-Poisson system, Hessian operator, Riemannian manifolds.

\section{Introduction}

Let $(M,g)$ be a finite dimensional Riemannian manifold and $X\in \mathcal{X}(M)$ a smooth vector field with $x_e$ an equilibrium point for the dynamics generated by the vector field $X$. Our interest in this paper is to give sufficient conditions for Lyapunov stability of the equilibrium point in some degenerate case that will be explained in the following. Typically, for degenerate cases we do not obtain stability with respect to all the variables of the manifold $M$, but for some cases we can obtain stability with respect to a part of the variables determined by an invariant submanifold. 

From the geometry of the problem a set of constraint functions $F_1,...,F_k:M\rightarrow \R$ is known. In the Hamilton-Poisson case the constraint functions are given by Casimir functions and sub-Casimir functions, see \cite{leonard-automatica}, \cite{leonard-1996}, \cite{leonard-1996-1}, \cite{leonard-marsden}. A nongeneric equilibrium point $x_e$ is a non-regular point for the function ${\bf F}=(F_1,...,F_k)$, i.e. ${\bf grad} F_1(x_e),...,{\bf grad} F_k(x_e)$ are linear dependent vectors in $T_{x_e}(M)$. In the context of Hamilton-Poisson systems nongeneric equilibriums are points that belong to singular symplectic leaves and their stability have been extensively studied in \cite{leonard-automatica}, \cite{leonard-marsden}, \cite{montaldi}, \cite{ortega-planas-ratiu}, \cite{ortega-1}, \cite{patrick-2003}, \cite{patrick-roberts}.  
Nevertheless, in some cases one can find a submanifold $x_e\in\widetilde{M}\subset M$ and a subset $F_{i_1},...,F_{i_q}$ of the constraint functions such that $x_e$ is a regular point for $\widetilde{F}_{i_1}:={F_{i_1}}_{|\widetilde{M}},...,\widetilde{F}_{i_q}:={F_{i_q}}_{|\widetilde{M}}:\widetilde{M}\rightarrow \R$. 

We introduce the submanifold $\widetilde{S}_e\subset \widetilde{M}$ as the preimage of the regular value $(\widetilde{F}_{i_1}(x_e),...,\widetilde{F}_{i_q}(x_e))\in \R^q$. We work in the hypothesis that $\widetilde{S}_e$ is invariant under the dynamics generated by $X$. The direct method of Lyapunov for the induced dynamics on $\widetilde{S}_e$ becomes: suppose there exists a smooth function $\widetilde{G}_e:\widetilde{S}_e\rightarrow \R$ such that:
\begin{itemize}
\item [(i)] $\dot{\widetilde{G}}_e:=d\widetilde{G}_e(X_{|\widetilde{S}_e})\leq 0$,

\item [(ii)] $d\widetilde{G}_e(x_e)=0$,

\item [(iii)] the Hessian matrix $\mathcal{H}^{\widetilde{G}_e}(x_e)$ is positive definite,
\end{itemize}
 then $x_e$ is a stable equilibrium point for the induced dynamics on the leaf $\widetilde{S}_e$.

A special case is when $\widetilde{M}$ is an invariant submanifold under the dynamics generated by the vector field $X$, $\widetilde{F}_{i_1},...,\widetilde{F}_{i_q}$ are conserved quantities for this induced dynamics, and $\widetilde{G}:\wt{M}\rightarrow \R$ is also a conserved quantity with the property that $\wt{G}_{|\wt{S}_e}=\wt{G}_e$. If moreover:
\begin{itemize}
\item [(a)] $d(\wt{G}_{|\wt{S}_e})(x_e)=0$,

\item [(b)]  the Hessian matrix $\mathcal{H}^{\widetilde{G}_{|\wt{S}_e}}(x_e)$ is positive definite,
\end{itemize}
 then $x_e$ is a stable equilibrium point for the induced dynamics on the leaf $\widetilde{S}_e$.

The passage from stability for dynamics induced on an invariant regular leaf $\wt{S}_e$ to the ambient space $\wt{M}$  is a consequence of Arnold method \cite{arnold}, energy-Casimir method \cite{holm-marsden-ratiu-weinstein}, Ortega-Ratiu method \cite{ortega-2}, algebraic method \cite{comanescu}, \cite{comanescu-1}, \cite{comanescu-2}.

The condition (a) is equivalent with the following equality, see \cite{birtea-comanescu-hess}, \cite{birtea-comanescu}:
\begin{itemize} 
\item [(a')] $\text{\bf grad }\wt{G}(x_e)=\sum_{s=1}^q\sigma_{i_s}(x_e)\text{\bf grad }\wt{F}_{i_s}(x_e)$,
\end{itemize}
where $\sigma_{i_1}(x_e),...,\sigma_{i_q}(x_e)$ are the Lagrange multipliers given by the formula 
\begin{equation}\label{sigma}
\sigma_{i_s}(x_e):=\frac{\det \Sigma_{(\wt{F}_{i_1},\ldots ,\wt{F}_{i_{s-1}},\wt{G}, \wt{F}_{i_{s+1}},\dots,\wt{F}_{i_q})}^{(\wt{F}_{i_1},\ldots ,\wt{F}_{i_{s}},\dots,\wt{F}_{i_q})}(x_e)}{\det \Sigma_{(\wt{F}_{i_1},\dots,\wt{F}_{i_q})}^{(\wt{F}_{i_1},\dots,\wt{F}_{i_q})}(x_e)},
\end{equation}
with the Gramian matrix defined by
\begin{equation}\label{sigma}
\Sigma_{(g_1,...,g_s)}^{(f_1,...,f_r)}=\left[%
\begin{array}{cccc}
  <\text{\bf grad } g_1,\text{\bf grad } f_{1}> & ... & <\text{\bf grad } g_s,\text{\bf grad } f_{1}> \\
  ... & ... & ... \\

  <\text{\bf grad } g_1,\text{\bf grad } f_r> & ... & <\text{\bf grad } g_s,\text{\bf grad } f_r> \\
\end{array}%
\right].
\end{equation}
The gradients in the above formulas are computed with respect to the induced metric on $\wt{M}$ by the ambient Riemannian space $(M,g)$. 

Further, the condition (b) is equivalent with the following equality, see \cite{birtea-comanescu-hess}:
\begin{itemize}
\item [(b')] $\left[ \mathcal{H}^{\wt{G}}(x_e)\right]_{|_{T_{_{x_e}}\wt{S}_e\times T_{_{x_e}}\wt{S}_e}}-\sum_{s=1}^q\sigma_{i_s}(x_e)\left[ \mathcal{H}^{\wt{F}_{i_s}}(x_e)\right]_{|_{T_{_{x_e}}\wt{S}_e\times T_{_{x_e}}\wt{S}_e}}$ is positive definite,
\end{itemize}
where Hessian matrices are also computed with respect to the induced metric on $\wt{M}$.

\begin{thm}
Suppose that $\wt{M}\subset M$ is an invariant submanifold under the dynamics, $\wt{F}_{i_1},...,\wt{F}_{i_q},\wt{G}$ are conserved quantities that satisfy conditions $(a')$ and $(b')$ then the nongeneric equilibrium point $x_e$ is Lyapunov stable on the invariant space $\wt{M}$.
\end{thm}

In general, conditions (a') and (b') does not imply stability in the whole ambient space $M$.

\section{Stability of nongeneric equilibria of underwater vehicles with noncoincident centers}

Following \cite{leonard-1996}, the dynamics of an underwater vehicle modeled as a neutrally buoyant, submerged rigid body in an infinitely large volume of irrotational, incompressible, inviscid fluid that is at rest at infinity is described by the system
\begin{equation}\label{underwater}
\left\{%
\begin{array}{ll}
\dot{\boldsymbol \Pi}={\boldsymbol \Pi}\times {\boldsymbol \Omega}+{\bf P}\times {\bf v}-mgl{\boldsymbol \Gamma}\times {\bf r} \\
\dot{\bf P}={\bf P}\times {\boldsymbol \Omega} \\
\dot{\boldsymbol \Gamma }={\boldsymbol \Gamma}\times {\boldsymbol \Omega},
\end{array}%
\right.
\end{equation}  
where ${\boldsymbol \Pi}$ is the angular impulse, ${\bf P}$ is the linear impulse, ${\boldsymbol \Gamma}$ is the direction of gravity, $l{\bf r}$ is the vector from center of buoyancy to center of gravity (with $l\geq 0$ and ${\bf r}$ a unit vector), $m$ is the mass of the vehicle, $g$ is gravitational acceleration, ${\boldsymbol \Omega}$ and ${\bf v}$ are the angular and translational velocity of the vehicle. In a body-fixed frame with the origin in the center of buoyancy the relationship between $({\boldsymbol \Pi},{\bf P})$ and  $({\boldsymbol \Omega},{\bf v})$ is given by
\begin{equation}\label{relation-between}
\left(%
\begin{array}{c}
{\boldsymbol \Pi} \\
{\bf P}
\end{array}%
\right)=\left(%
\begin{array}{cc}
J & D \\
D^T & M
\end{array}%
\right)\left(%
\begin{array}{c}
{\boldsymbol \Omega} \\
{\bf v}
\end{array}%
\right)
,
\end{equation}
where $J$ is the matrix that is the sum of the body inertia matrix plus the added inertia matrix associated with the potential flow model of the fluid, $M$ is the sum of the mass matrix for the body alone, and $D$ accounts for the cross terms. 
The relationship between $({\boldsymbol \Omega},{\bf v})$ and $({\boldsymbol \Pi},{\bf P})$ is given by
\begin{equation}\label{relation-between}
\left(%
\begin{array}{c}
{\boldsymbol \Omega} \\
{\bf v}
\end{array}%
\right)=
\left(%
\begin{array}{cc}
A & B^T \\
B & C
\end{array}%
\right)
\left(%
\begin{array}{c}
{\boldsymbol \Pi} \\
{\bf P}
\end{array}%
\right),\,\,\left(%
\begin{array}{cc}
A & B^T \\
B & C
\end{array}%
\right)=\left(%
\begin{array}{cc}
J & D \\
D^T & M
\end{array}%
\right)^{-1}.
\end{equation}
In this paper we suppose that
\begin{equation}
M=\left(%
\begin{array}{ccc}
m_1 & 0 & 0 \\
0 & m_2 & 0\\
0 & 0 & m_{3}
\end{array}%
\right),
J=\left(%
\begin{array}{ccc}
I_{1} & 0 & 0 \\
0 & I_{2} & 0\\
0 & 0 & I_{3}
\end{array}%
\right)
,
D=ml\widehat{\bf r}=\left(%
\begin{array}{ccc}
0 & -ml & 0 \\
ml & 0 & 0\\
0 & 0 & 0
\end{array}%
\right).
\end{equation}
By a direct computation we obtain:
$$A=\left(%
\begin{array}{ccc}
a_1 & 0 & 0 \\
0 & a_2 & 0\\
0 & 0 & \frac{1}{I_3}
\end{array}%
\right),\,\,B=\left(%
\begin{array}{ccc}
0 & b_1 & 0  \\
-b_2 & 0 & 0 \\
 0 & 0 & 0
\end{array}%
\right),\,\,C=\left(%
\begin{array}{ccc}
c_1 & 0 & 0 \\
0 & c_2 & 0\\
0 & 0 & \frac{1}{m_3}
\end{array}%
\right),$$
where
$$a_1=\frac{m_2}{m_2I_1-m^2l^2},\,\, a_2=\frac{m_1}{m_1I_2-m^2l^2},\,\, b_1=-\frac{ml}{m_1I_2-m^2l^2},$$ 
$$ b_2=-\frac{ml}{m_2I_1-m^2l^2},\,\, c_1=\frac{I_2}{m_1I_2-m^2l^2},\,\, c_2=\frac{I_1}{m_2I_1-m^2l^2}.$$
From physical reasoning we have $a_1>0,a_2>0,c_1>0,c_2>0$ and $b_1<0,b_2<0$, see \cite{leonard-automatica}.

The system \eqref{underwater} has the Hamilton-Poisson formulation on the Poisson manifold $M=\frak{se}^*(3)$, see \cite{leonard-automatica}, \cite{leonard-marsden}:
\begin{equation}\label{underwater-1-Poisson}
\dot{\bf z}=\Lambda({\bf z})\nabla H({\bf z}),
\end{equation}
where ${\bf z}=({\boldsymbol \Pi},{\bf P},{\boldsymbol \Gamma})$, 
$\Lambda({\bf z})=\left(%
\begin{array}{ccc}
\widehat{{\boldsymbol \Pi}} & \widehat{\bf P} & \widehat{{\boldsymbol \Gamma}} \\
\widehat{\bf P} & O_3 &O_3 \\
 \widehat{{\boldsymbol \Gamma}} & O_3 &O_3 
\end{array}%
\right),$ and the Hamiltonian function is given by
$$H({\bf z})=\frac{1}{2}(<{\boldsymbol \Pi},A{\boldsymbol \Pi}>+2<{\boldsymbol \Pi},B^T{\bf P}>+<{\bf P},C{\bf P}>-2mgl<{\boldsymbol \Gamma},{\bf r}>).
$$
The Poisson structure induced by $\Lambda$ has the following Casimir functions:
$$C_1({\bf z})=<{\bf P},{\boldsymbol \Gamma}>, \,\, C_2({\bf z})=\frac{1}{2}||{\bf P}||^2,\,\,C_3({\bf z})=\frac{1}{2}||{\boldsymbol \Gamma}||^2.$$
We have a set of sub-Casimir functions:
$$C_4({\bf z})=<{\boldsymbol \Pi},{\bf P}>,\,\,C_5({\bf z})=<{\boldsymbol \Pi},{\boldsymbol \Gamma}>.$$
The set of interest is given by the following nongeneric equilibria with spin:
$$\mathcal{E}=\left\{{\bf z}_e=({\boldsymbol \Pi}_e,{\bf P}_e,{\boldsymbol \Gamma}_e)\,|\,{\boldsymbol \Pi}_e=
\left(%
\begin{array}{c}
 0 \\
 0\\
\Pi_e
\end{array}%
\right),
{\bf P}_e=
\left(%
\begin{array}{c}
 0 \\
 0\\
P_e
\end{array}%
\right),
{\boldsymbol \Gamma}_e=
\left(%
\begin{array}{c}
 0 \\
 0\\
1
\end{array}%
\right),\,\,\Pi_e\neq 0\right\}.$$
An equilibrium point ${\bf z}_e\in \mathcal{E}$ is nongeneric because it does not belong to a regular symplectic leaf as $\nabla C_1({\bf z}_e), \nabla C_2({\bf z}_e), \nabla C_3({\bf z}_e)$ are linear dependent vectors. From the set of constraint functions $C_1,C_2$, $C_3$, $C_4,C_5$ we can choose $C_1,C_3,C_5$ such that ${\bf z}_e$ belongs to the regular leaf $$L_e:=(C_1,C_3,C_5)^{-1}(C_1({\bf z}_e),C_3({\bf z}_e),C_5({\bf z}_e)).$$
We cannot apply yet  the stability result from the previous section as the leaf $L_e$ is not an invariant set under the dynamics \eqref{underwater}.

Because $C_1,C_2,C_3$ are conserved quantities we obtain that the set 
$$\mathcal{M}=\{{\bf z}=({\boldsymbol \Pi},{\bf P},{\boldsymbol \Gamma})\,|\,\nabla C_1({\bf z}),\nabla C_2({\bf z}),\nabla C_3({\bf z}) \,\,\text{are linear dependent vectors}\}$$
is an invariant set for \eqref{underwater}, see \cite{birtea-comanescu-invariant}. By a direct computation we obtain that 
$\mathcal{M}=\{{\bf z}=({\boldsymbol \Pi},{\bf P},{\boldsymbol \Gamma})\,|\,{\bf P}\parallel {\boldsymbol \Gamma}\}$, which from a Poisson manifold point of view is the manifold formed by all degenerate symplectic leaves.

We consider the invariant set $\wt{M}\subset \frak{se}^*(3)$, 
$$\wt{M}=\{{\bf z}=({\boldsymbol \Pi},{\bf P},{\boldsymbol \Gamma})\,|\,{\bf P}\parallel {\boldsymbol \Gamma}\,\text{with}\,{\bf P}\neq {\bf 0}\,\,\text{or}\,\,{\boldsymbol \Gamma}\neq {\bf 0}\},$$
which is the union of all degenerate symplectic leaves of dimension 4. The set $\wt{M}$ has a structure of a 7-dimensional manifold with an atlas which contains two local charts, $\Phi_{\Gamma},\Phi_P:\R^3\times\R^3\backslash\{{\bf 0}\}\times \R\rightarrow \wt{M}$,
\begin{align*}
\Phi_{\Gamma}(x_1,x_2,x_3,x_4,x_5,x_6,x_7) & :=(x_1,x_2,x_3,x_4x_7,x_5x_7,x_6x_7,x_4,x_5,x_6), \\
\Phi_{P}(y_1,y_2,y_3,y_4,y_5,y_6,y_7) & :=(y_1,y_2,y_3,y_4,y_5,y_6,y_4y_7,y_5y_7,y_6y_7).
\end{align*}
The two local charts are compatible with the coordinate transformation $$\phi_{\Gamma P}:=\Phi_P^{-1}\circ\Phi_{\Gamma}:\R^3\times\R^3\backslash\{{\bf 0}\}\times \R\backslash\{0\}\rightarrow \R^3\times\R^3\backslash\{{\bf 0}\}\times \R\backslash\{0\},$$
$$\phi_{\Gamma P}(x_1,x_2,x_3,x_4,x_5,x_6,x_7)=(x_1,x_2,x_3,x_4x_7,x_5x_7,x_6x_7,\frac{1}{x_7}).$$

For the induced dynamics on the invariant manifold $\wt{M}$ we have the conserved quantities $\wt{C}_1,\wt{C}_2,\wt{C}_3,\wt{C}_4$,  
 $\wt{C}_5:\wt{M}\rightarrow \R$, $\wt{C}_i:={C_i}_{|\wt{M}}$, $i\in \overline{1,5}$. The one forms $d\wt{C}_1({\bf z}_e),...,d\wt{C}_5({\bf z}_e)$ are linear dependent one forms, where ${\bf z}_e\in \mathcal{E}$. Choosing an equilibrium point ${\bf z}_e\in \mathcal{E}$ and $\wt{C}_1,\wt{C}_3,\wt{C}_5$, we define the regular leaf $\wt{S}_e:=(\wt{C}_1,\wt{C}_3,\wt{C}_5)^{-1}(\wt{C}_1({\bf z}_e),\wt{C}_3({\bf z}_e),\wt{C}_5({\bf z}_e))\subset \wt{M}$, which is an invariant submanifold under the dynamics as $\tilde{C}_5$ becomes a conserved quantity for the induced dynamics on $\wt{M}$.

The induced Riemannian metric $\wt{g}$ on the submanifold $\wt{M}$ by the Euclidean metric on $\R^9$ in the local chart $\Phi_{\Gamma}$ has the associated matrix 
$$\left[ \wt{g}\right]=\left(%
\begin{array}{ccccccc}
 1 & 0 & 0 & 0 & 0 & 0 & 0 \\
  0 & 1 & 0 & 0 & 0 & 0 & 0 \\
 0 & 0 & 1 & 0 & 0 & 0 & 0 \\
 0 & 0 & 0 & x_7^2+1 & 0 & 0 & x_4x_7 \\
 0 & 0 & 0 & 0 &  x_7^2+1 & 0 & x_5x_7 \\
 0 & 0 & 0 & 0 &  0 & x_7^2+1 & x_6x_7 \\
0 & 0 & 0 & x_4x_7 &  x_5x_7 & x_6x_7 & x_4^2+x_5^2+x_6^2 
\end{array}%
\right). $$

For computational reasoning we consider the symmetric case $I_1=I_2$ and $m_1=m_2$ which imply $a_1=a_2\stackrel{not}{=}a$,   $b_1=b_2\stackrel{not}{=}b$, and  $c_1=c_2\stackrel{not}{=}c$. According to \cite{leonard-automatica}, \cite{leonard-marsden} we obtain the supplementary conserved quantity $K=\Pi_3$ for the system \eqref{underwater}. As before, we make the notation $\wt{K}=K_{|\wt{M}}$.

Searching for stable equilibria it is sufficient to find $\lambda\in \R$ such that the conserved quantity $\wt{G}_{\lambda}:=\wt{H}+\lambda\wt{K}$ verifies the conditions (a') and (b') described in the above section. All the equilibrium points in $\mathcal{E}$ belong to the domain of the local chart $\Phi_{\Gamma}$ and in this chart they have the coordinates:
$$\mathcal{E}=\{(0,0,\Pi_e,0,0,1,P_e)\,|\,\Pi_e\neq 0\}.$$
In the local chart $\Phi_{\Gamma}$ we also have:
\begin{align*}
\wt{H} & =\frac{a}{2}x_1^2+\frac{a}{2}x_2^2+\frac{1}{2I_3}x_3^2+bx_1x_5x_7-bx_2x_4x_7+\frac{c}{2}x_4^2x_7^2+\frac{c}{2}x_5^2x_7^2+\frac{1}{2m_3}x_6^2x_7^2-mglx_6, \\
\wt{C}_1 & =x_4^2x_7+x_5^2x_7+x_6^2x_7, \\
\wt{C}_3 & =\frac{1}{2}(x_4^2+x_5^2+x_6^2), \\
\wt{C}_5 & =x_1x_4+x_2x_5+x_3x_6, \\
\wt{K} & =x_3.
\end{align*}

Using the induced metric on $\wt{M}$ and formulas \eqref{sigma}, we obtain the Lagrange multipliers at an equilibrium point ${\bf z}_e\in \mathcal{E}$ for the function $\wt{G}_{\lambda}$:
\begin{align*}
\sigma_{\wt{C}_1}({\bf z}_e) & =\frac{P_e}{m_3}, \\
\sigma_{\wt{C}_3}({\bf z}_e) & =-\frac{I_3P_e^2+m_3\Pi_e^2+\lambda I_3 m_3 \Pi_e+mgl I_3m_3}{I_3m_3}, \\
\sigma_{\wt{C}_5}({\bf z}_e) & =\frac{\lambda I_3+\Pi_e}{I_3}.
\end{align*}

By a straight forward computation we obtain that condition (a') is verified for any equilibrium point ${\bf z}_e\in \mathcal{E}$.

We are searching for equilibrium points ${\bf z}_e\in \mathcal{E}$ such that the Hessian matrix of the function $\wt{G}_{\lambda}$ restricted to the leaf $\wt{S}_e$ is positive definite; i.e. $\mathcal{H}^{\wt{G}_{{\lambda}_{|\wt{S}_e}}}({\bf z}_e)>0$ (condition (b')). Equivalently,  
\begin{equation}\label{inequality}
\left(\left[ \mathcal{H}^{\wt{G}_{\lambda}}({\bf z}_e)\right]-\sigma_{\wt{C}_1}({\bf z}_e)\left[ \mathcal{H}^{\wt{C}_1}({\bf z}_e)\right]-\sigma_{\wt{C}_3}({\bf z}_e)\left[ \mathcal{H}^{\wt{C}_3}({\bf z}_e)\right]-\sigma_{\wt{C}_5}({\bf z}_e)\left[ \mathcal{H}^{\wt{C}_5}({\bf z}_e)\right]\right)_{|_{T_{_{{\bf z}_e}}\wt{S}_e\times T_{_{{\bf z}_e}}\wt{S}_e}}>0.
\end{equation}
The Hessian matrix of a function $\wt{F}:(\wt{M},\wt{g})\rightarrow \R$  is computed with the formula 
$$\mathcal{H}^{\wt{F}}_{ij}({\bf z}_e)=\frac{\partial^2\wt{F}}{\partial x^{i}\partial x^{i}}({\bf z}_e)-\Gamma_{ij}^{k}({\bf z}_e)\frac{\partial \wt{F}}{\partial x^{k}}({\bf z}_e),\,\,i,j,k=\overline{1,7}.$$
The Christoffel's symbols associated to the induced metric $\wt{g}$ are:
$$\Gamma^4_{47}({\bf z}_e)=\Gamma^4_{74}({\bf z}_e)=\Gamma^5_{57}({\bf z}_e)=\Gamma^5_{75}({\bf z}_e)=\frac{P_e}{P_e^2+1},\,\,\Gamma^7_{67}({\bf z}_e)=\Gamma^7_{76}({\bf z}_e)=1,\,\,\text{all the rest being zero}.$$
For the tangent space to the leaf $\wt{S}_e$ in the equilibrium point ${\bf z}_e$ we choose the base:
$${\bf w}_1=(1,0,0,0,0,0,0),\,\,{\bf w}_2=(0,1,0,0,0,0,0),\,\,{\bf w}_3=(0,0,0,1,0,0,0),\,\,{\bf w}_4=(0,0,0,0,1,0,0).$$
The left hand side of \eqref{inequality} is the $4\times 4$ symmetric matrix with the following entries:
\begin{align*}
h_{{11}} & ={\frac {m_{{1}}}{m_{{1}}I_{{1}}-{m}^{2}{l}^{2}}},\,\,h_{12}=0,\,\,h_{{13}}=-{\frac {\lambda\,I_{{3}}+\Pi _{{e}}}{I_{{3}}}},\,\,h_{{14}}={\frac {mlP_{{e}}}{m_{{1}}I_{{1}}-{m}^{2}{l}^{2}}},\\
h_{{22}} & ={\frac {m_{{1}}}{m_{{1}}I_{{1}}-{m}^{2}{l}^{2}}},\,\,h_{{23}}=-{\frac {mlP_{{e}}}{m_{{1}}I_{{1}}-{m}^{2}{l}^{2}}},\,\,h_{{24}}=-{\frac {\lambda\,I_{{3}}+\Pi _{{e}}}{I_{{3}}}},\\
h_{{33}} & ={\frac {{P_{{e}}}^{2}I_{{1}}m_{{3}}I_{{3}}+{m}^{2}{l}^{2}{P_{
{e}}}^{2}I_{{3}}-m_{{1}}I_{{1}}{P_{{e}}}^{2}I_{{3}}-I_{{3}}g{m}^{3}{l}
^{3}m_{{3}}-\Pi _{{e}}m_{{3}}\lambda\,I_{{3}}{m}^{2}{l}^{2}+\Pi _{{e}}
m_{{3}}\lambda\,I_{{3}}m_{{1}}I_{{1}}}{ \left( m_{{1}}I_{{1}}-{m}^{2}{l}^{2} \right) I_{{3}}m_{{3}}}} \\
& +\frac{mglm_{{3}}m_{{1}}I_{{1}}I_{{3}}-
{\Pi _{{e}}}^{2}m_{{3}}{m}^{2}{l}^{2}+{\Pi _{{e}}}^{2}m_{{3}}m_{{1}}I_
{{1}}}{ \left( m_{{1}}I_{{1}}-{m}^{2}{l}^{2} \right) I_{{3}}m_{{3}}},\,\,h_{34}=0, \\
h_{{44}} & ={\frac {{P_{{e}}}^{2}I_{{1}}m_{{3}}I_{{3}}+{m}^{2}{l}^{2}{P_{
{e}}}^{2}I_{{3}}-m_{{1}}I_{{1}}{P_{{e}}}^{2}I_{{3}}-I_{{3}}g{m}^{3}{l}
^{3}m_{{3}}-\Pi _{{e}}m_{{3}}\lambda\,I_{{3}}{m}^{2}{l}^{2}+\Pi _{{e}}
m_{{3}}\lambda\,I_{{3}}m_{{1}}I_{{1}}}{\left( m_{{1}}I_{{1}}-{m}^{2}{l}^{2} \right) I_{{3}}m_{{3}}}} \\
& +{\frac{mglm_{{3}}m_{{1}}I_{{1}}I_{{3}}-
{\Pi _{{e}}}^{2}m_{{3}}{m}^{2}{l}^{2}+{\Pi _{{e}}}^{2}m_{{3}}m_{{1}}I_
{{1}}}{ \left( m_{{1}}I_{{1}}-{m}^{2}{l}^{2} \right) I_{{3}}m_{{3}}}}.
\end{align*}
To study the positive definiteness of the above matrix we apply the Sylvestre's criterion. The first principal minor equals $a$ which is positive from physical reasoning. The second principal minor equals $a^2$. The third principal minor (we denote by $\Theta_3$) and fourth principal minor (we denote it by $\Theta_4$) are related by the equality $\Theta_4=\frac{1}{a^2}\Theta_3^2$. Consequently, the reduced Hessian matrix is positive definite if and only if $\Theta_3>0$. We have the expression:
\begin{align*}
\Theta_3 & =-\frac{a^2}{m_1m_3I_3^2}\left( m_{{3}}{I_{{3}}}^{2}(m_1I_1-{m}^{2}{l}^{2}\right) {\lambda}^{2}+ \left( 2\,\Pi _{{e}}m_{{3}}I_{{3}}m_{{
1}}I_{{1}}-m_{{1}}\Pi _{{e}}m_{{3}}{I_{{3}}}^{2}-2\,\Pi _{{e}}m_{{3}}I
_{{3}}{m}^{2}{l}^{2} \right) \lambda \\
 & \quad+{\Pi _{{e}}}^{2}m_{{3}}m_{{1}}I_{
{1}}+m_{{1}}{P_{{e}}}^{2}{I_{{3}}}^{2}-m_{{1}}{I_{{3}}}^{2}{\it mgl}\,
m_{{3}}-m_{{1}}I_{{3}}{\Pi _{{e}}}^{2}m_{{3}}-{\Pi _{{e}}}^{2}m_{{3}}{
m}^{2}{l}^{2}-{P_{{e}}}^{2}m_{{3}}{I_{{3}}}^{2})
\end{align*}
A straight forward analysis shows that there exists $\lambda\in \R$ such that $\Theta_3>0$ if and only if 
\begin{equation}\label{conditie-stabilitate}
mgl>\left(\frac{1}{m_3}-\frac{1}{m_1}\right)P_e^2-\frac{a}{4}\Pi_e^2.
\end{equation}
The above inequality is the same condition obtained in \cite{leonard-automatica}, \cite{leonard-marsden} and it ensures the stability for the nongeneric equilibrium point ${\bf z}_e$ with respect to the  induced dynamics on the leaf $\wt{S}_e$. Thus, by Arnold method \cite{arnold}, energy-Casimir method \cite{holm-marsden-ratiu-weinstein}, Ortega-Ratiu method \cite{ortega-2}, algebraic method \cite{comanescu}, \cite{comanescu-1}, \cite{comanescu-2} we obtain the stability with respect to the induced dynamics on $\wt{M}$.

It has been proved in \cite{leonard-marsden}, \cite{leonard-automatica} that condition 
$$\left(\frac{1}{m_3}-\frac{1}{m_1}\right)P_e^2<mgl$$
implies nonlinear stability on the whole space $\frak{se}^*(3)$ by using a Lyapunov function that does not imply sub-Casimir functions but only the Hamiltonian function and Casimir functions that are conserved quantities for \eqref{underwater}. It also has been proved in the same papers that condition 
\begin{equation*}
\left(\frac{1}{m_3}-\frac{1}{m_1}\right)P_e^2-\frac{a}{4}\Pi_e^2<mgl<\left(\frac{1}{m_3}-\frac{1}{m_1}\right)P_e^2
\end{equation*}
implies spectral stability. For proving nonlinear stability in this case one needs to take into account the sub-Casimir functions that are not conserved quantities for the dynamics on the whole space. Special care has to be considered in this case.

\begin{thm} Let  ${\bf z}_e\in \mathcal{E}$ be a nongeneric equilibrium with spin. We have the following nonlinear stability behavior:
\begin{itemize}
\item [(i)] If $\left(\frac{1}{m_3}-\frac{1}{m_1}\right)P_e^2<mgl$ then the equilibrium point is Lyapunov stable on the whole space $\frak{se}^*(3)$.
\item [(ii)] If $\left(\frac{1}{m_3}-\frac{1}{m_1}\right)P_e^2-\frac{a}{4}\Pi_e^2<mgl\leq\left(\frac{1}{m_3}-\frac{1}{m_1}\right)P_e^2$ then the equilibrium point is Lyapunov stable for the induced dynamics on the invariant submanifold $\wt{M}$.
\item [(iii)] If $mgl< \left(\frac{1}{m_3}-\frac{1}{m_1}\right)P_e^2-\frac{a}{4}\Pi_e^2$ then the equilibrium point is unstable.
\end{itemize}
\end{thm}

It remains to be studied if the inequality $(ii)$ in the above theorem also guaranties the stability of a nongeneric equilibrium ${\bf z}_e\in \mathcal{E}$ with respect to the whole space $\frak{se}^*(3)$.

\bigskip

{\bf Acknowledgments.} This work was supported by a grant of the Romanian National Authority for Scientific Research, CNCS - UEFISCDI, under the Romania-Cyprus bilateral cooperation programme  (module III), project number 760/2014.  We are thankful to Ioan Casu for his help with Maple programming.

\end{document}